%% file: main.tex
\def\year{2021}\relax
\documentclass[letterpaper, twocolumn]{article} 
\usepackage{aaai21}  
\usepackage{times}  
\usepackage{helvet} 
\usepackage{courier}  
\usepackage[hyphens]{url}  
\usepackage{graphicx} 
\usepackage{natbib}  
\usepackage{caption} 
\usepackage{amsmath}

\usepackage{romannum}
\usepackage{amssymb}

\usepackage{booktabs}
\usepackage[table]{xcolor}
\definecolor{grey}{rgb}{0.8,0.8,0.8}
\definecolor{lgrey}{rgb}{0.9,0.9,0.9}
\usepackage[labelformat=simple]{subcaption}

\usepackage{algorithm}
\usepackage{algpseudocode}

\usepackage{amsmath} 
\usepackage[acronym, toc, shortcuts, nowarn]{glossaries}

\title{Robust Multi-Agent Reinforcement Learning with Social Empowerment for Coordination and Communication}

\author {
        Tessa van der Heiden,\textsuperscript{\rm 1}
        Herke van Hoof, \textsuperscript{\rm 2}
        Efstratios Gavves, \textsuperscript{\rm 2}
        Christoph Salge \textsuperscript{\rm 3} \\
}
\affiliations {
    \textsuperscript{\rm 1} BMW Group \\
    \textsuperscript{\rm 2} University of Amsterdam \\
    \textsuperscript{\rm 3} University of Hertfordshire \\
    tessa.heiden@bmw.de, h.c.vanhoof@uva.nl, 
    egavves@uva.nl, c.salge@herts.ac.uk
}

\newacronym{RL}{RL}{Reinforcement learning}
\newacronym{MARL}{MARL}{Multi-agent reinforcement learning}
\newacronym{SP}{SP}{Self-play}
\newacronym{IM}{IM}{Intrinsic motivation}
\usepackage[switch]{lineno} 
\begin{document}

\maketitle

\begin{abstract}
We consider the problem of robust multi-agent reinforcement learning (MARL) for cooperative communication and coordination tasks. MARL agents, mainly those trained in a centralized way, can be brittle because they can overfit to their training partners. This overfitting can produce agents that adopt policies that act under the expectation that other agents will act a certain way rather than react to their actions. Our objective is to bias the learning process towards finding strategies that remain reactive towards others' behavior. 
Social empowerment measures the \textit{potential} influence between agents' actions.
We propose it as an additional reward term, so agents better adapt to other agents' actions.
 We show that the proposed method results in obtaining higher rewards faster and a higher success rate in three cooperative communication and coordination tasks.
\end{abstract}
 
\input{introduction}

\input{related_work}

\input{preliminaries}
\input{methodology}

\input{experiments}

\input{conclusion}

\bibliography{bib}
\end{document}

%% file: introduction.tex
\section{Introduction}\label{se: intro}
\ac{MARL} holds considerable promise to help address a variety of cooperative multi-agent problems. However, a problem with \ac{MARL} is that agents can derive strong policies that are overfitted to their partners' behavior. Specifically, with centralized training agents can adopt strategies that expect other agents to act in a certain way rather than react to their actions. Such systems are undesirable as they may fail when their partners alter their strategies or have to collaborate with novel partners, either during the learning or deployment phase. 

Robust \ac{RL} was introduced by \citet{morimoto2005robust} considering the generalization ability of the learned policy in the single-agent setting. 
We consider robustness to be a training agent's ability to deal with changes in the other agents' behavior.

Our objective is to obtain multi-agent policies that avoid this specific lack of robustness and stay reactive to changes in other agents' policy. 
Agents acting without any sensor input and attention towards their partners cannot have this ability.
We want to introduce an additional reward to bias learning towards socially reactive strategies - which should fulfil the following constraints: 1) it should, with minimal adaptation, apply to a wide range of problems with different sensor-actuator configurations, to preserve the universality of the RL framework, 2) it should not negatively affect the performance, i.e., once good policies are found, it should not harm exploitation. Optimizing for the mutual information between agents' actions \cite{jaques2019social}, requires the agents' policies to have a certain degree of entropy and this might interfere with an exploitation strategy. Fulfilling the above criteria would provide a general-purpose multi-agent learning algorithm for various cooperative tasks.

%% file: related_work.tex
\section{Related work}

\subsection{Robust Multi-Agent Reinforcement Learning}
%
%

There is a large body of research on constructing agents that are robust to their partners. In self-play, for example, agents train against themselves rather than a fixed opponent strategy to prevent developing exploitable strategies \cite{tesauro1994td}. Population based-training goes one step further by training agents to play against a population of other agents rather than only a copy of itself. For instance,  \citet{max2018human} and \citet{lowe2017multi} train an ensemble of policies with a variety of collaborators and competitors. 
By using a whole population rather than only a copy of itself, the agent is forced to be robust to a wide variety of potential perturbations instead of a single perturbation.
However, it requires a great deal of engineering because the policy parameters suitable for the previous environment are not necessarily the next stage's best initialization. 

Some works combine the minimax framework and \ac{MARL} to find policies that are robust to opponents with different strategies. Minimax is a concept in game theory that can be applied to find an approach that minimizes the possible loss in a worst-case scenario \cite{osborne2004introduction}. \citet{li2019robust} use it during training to optimize the reward for each agent under the assumption that all other agents act adversarial. We are interested in methods that can deal with perturbations in the training partners' behavior, which differs from dealing with partners with various strategies. 

In zero-shot coordination, agents need to cooperate with novel partners that are not seen during training. Other-play \cite{hu2020other} constructs zero-shot coordination strategies by making their agents robust to how their partners break symmetries in the underlying system. 
We instead frame our problem as one-shot adaptation, i.e. our policy should be able to quickly adapt if the other agents alter their policies.


%
%
\subsection{Intrinsic Social Motivation}
Due to centralized training in MARL, agents might adopt non-reactive strategies that may struggle with other agents' changing behaviors. Social intrinsic motivation can give an additional incentive to find reactive policies towards other agents.

\ac{IM} in \ac{RL} refers to reward functions that allow agents to learn interesting behavior, sometimes in the absence of an environmental reward \cite{chentanez2005intrinsically}. 
Computational models of \ac{IM} are generally separated into two categories \cite{santucci2012intrinsic}, \cite{baldassarre2013intrinsically}, those that focus on curiosity \cite{burda2018large}, \cite{pathak2017curiosity} and exploration \cite{gregor2016variational}, \cite{eysenbach2018diversity}, and those that focus on competence and control \cite{oudeyer2009intrinsic}, \cite{karl2017unsupervised}. The information-theoretic Empowerment formalism is in the latter category, trying to capture how much an agent is in control of the world it can perceive. Empowerment has produced robust behaviour linked to controllability, operationality and self-preservation - in both robots \cite{van2020social}, \cite{karl2017unsupervised}, \cite{leu2013corbys} and simulations \cite{guckelsberger2016intrinsically}, with \cite{de2018unified} and without \cite{guckelsberger2018new} reinforcement learning and neural network approximations \cite{karl2017unsupervised}.

Empowerment has also been applied to multi-agent simulations, under the term of coupled empowerment maximization \cite{guckelsberger2016intrinsically}, in which it was used to produce supportive and antagonistic behavior. Of particular interest is the idea of transfer empowerment - a measure that quantifies behaviors such as collaboration, coordination, and lead-taking \cite{salge2017empowerment}. Empowerment has also been tested in collaborative scenarios involving humans and robots. In all scenarios, transfer empowerment provides an additional incentive for agents to interact and coordinate. 

There are similar techniques that quantify the interaction between agents for improving coordination between agents. \citet{barton2018measuring} analyze the degree of dependence between two agents’ policies to measure coordination, specifically by using Convergence Cross Mapping (CCM).  
\citet{strouse2018learning} show how agents can share (or hide) intentions by maximising the mutual information between actions and a categorical goal. 
One notably relevant work is by \citet{jaques2019social} that design social influence which is the influence of one agent on the policies of other agents, measured by the mutual information between action pairs of distinct agents. 
In contrast to social influence, transfer empowerment considers the \textit{potential} mutual information or channel capacity. When optimizing for \textit{actual} mutual information, then its value is bounded from above by the lowest entropy of both agent's action variables. This might easily interfere with an exploitation strategy and may need to be regularized once a good strategy is found. On the other hand, empowerment does not have this limitation and the action sets could have very narrow distributions.

%% file: preliminaries.tex
\section{Preliminaries}
Let us consider a Dec-POMDP, an extension of the MDP for multi-agent systems, being both decentralized and partially observable (Nair et al., 2003). This means that each of the $n$ agents conditions the choice of its action on its partial observation of the world. It is defined by the following tuple: $(\textbf{S}, \boldsymbol{A}, T, \boldsymbol{O}, O, R, n)$. $ \boldsymbol{S}$ is the set of true global states, while $\boldsymbol{O}=\otimes_{i \in [1, .., n]} \boldsymbol{O}_i$ is the set of joint local observations and $\boldsymbol{A}=\otimes_{i \in [1, .., n]} \boldsymbol{A}_i$ the set of possible joint actions. At each time step, the state transition function $T$ maps the joint action and state to a new state, with probability distribution $P(s_{t+1}|s_t, \boldsymbol{a}_t)$. 

We consider fully cooperative tasks, so agents share a reward $r_t$ which is conditioned on the joint action and state. $R$ is the reward function that maps the state and joint actions into a single scalar reward. 

%% file: methodology.tex
\section{Methodology}
In the introduction, we have argued that agents that are trained in a centralized manner can adopt brittle policies and perform poorly, even in simple coordination tasks. The reason is that they adopt strategies that expect other agents will act in a certain way rather than observe and anticipate their actions.

This section aims to describe an additional heuristic that biases the learning process in obtaining policies towards those that are reactive towards other agents' actions.

First we introduce our specific version of transfer empowerment, that rewards the idea of an agent being responsive to adaptations in the other's policy. Transfer empowerment is computed for a each pair of agents, which scales quadratically with the number of agents.

By adopting the framework of centralized training, we can compute empowerment over the joint action and states. This notion captures the causal influence of all agents' actions on the overall resulting state. Since it is calculated over the joint actions and states and not per pair, we arrive at a more scalable solution.

\subsection{Transfer Empowerment} \label{se: transfer}
Suppose there are two agents, $j$ and $k$, both taking actions and changing the overall state. Each time agent $k$ takes an action, the state of agent $j$ is modified and therefore agent $j$'s policy conditions on it. The objective of coordination is that by modifying the actions of agent $k$ , agent $j$ also \emph{reliably} adapts its actions. Here we look at transfer empowerment, namely the \emph{potential} causal influence that one agent has on another. It is computed for pairs of agents by maximizing the mutual information $\mathcal{I}$ between one agent’s action $a^k_t$ and another agent's action  $a^j_{t+1}$ at a later point in time, conditioned on the current and next state, $s_t$ and $s_{t+1}$ respectively:
\begin{equation}
    \mathcal{E}^{T, k\rightarrow j}(s_t) = \underset{\omega^k}{\max~} \mathcal{I}
    \left[a^j_{t+1}, a^k_t | s_t \right]
\end{equation}

Here, $\omega^k(a^k_{t+1} |s_{t+1})$ is the \emph{hypothetical} policy of agent $k$, that takes an action $a^k_{t+1}$ after observing state $s_{t+1}$ and influencing $a^{j}_{t+2}$ at a later time step. 

Our version of transfer empowerment differs slightly from the one introduced by \cite{salge2017empowerment}, as we consider the potential information flow, or channel capacity, between one agent’s action and another agent's action at a later point in time. 
 \citet{salge2017empowerment} consider the empowerment between one agent's action and another agent's \emph{sensor} state.
Transfer empowerment to another agent’s sensor captures the direct influence to change the other agents environment, and also influencing the agent to take certain actions, which then influences their world. Using transfer empowerment to another agent’s \emph{action} focuses on just the influence that affects the other agent's decision. Since an agent is usually in control of their actions, influence on their action has to flow through their sensor and be mediated by their policy. In other words, they have to react to the first agent’s actions. This is more in line with our goal of biasing a policy towards more reactive strategies.


Transfer empowerment has ties with but is different from social influence \cite{jaques2019social}. Social influence is the mutual information between agents' actions. It  is high when both action variables have a certain entropy, i.e., that different actions are performed. Once an acceptable policy is found, e.g. towards end of training, a high entropy policy distribution might actually be suboptimal. Empowerment on the other hand, considers the \textit{potential} and not \textit{actual }information flow, so the agent will only calculate how it could have affected the other agent, rather than actually carry out its potential. As such, action sets can have very narrow distributions, as long as the system would still be reactive \textit{if} those actions change. Therefore it does not interfere with the constraints stated earlier. 

Here we described transfer empowerment computed from agent $k$'s actions to agent's $j$' action, denoted by $\mathcal{E}^{T, k\rightarrow j}$. 
  $\mathcal{E}^{T, 1:n\rightarrow j} = \sum_{\substack{i =1 \\ i\neq j}}^n \mathcal{E}^{T, i\rightarrow j}$ is the extension for multiple agents and results in agent $j$ being reactive to agents' actions. The main goal is to train a  \emph{behavioral} policy $\pi^{j}$ to maximize the transfer empowerment, which is done by estimating $\mathcal{E}^T$ with $\omega^k$ and by changing $\pi^j$.


\subsection{Joint Empowerment as Scalable Proxy}
Ideally, we would want to compute the transfer empowerment from all agents to all other agents. However, this grows quadratically and might become quickly infeasible. Therefore we introduce joint empowerment as a better computable heuristic proxy. 
We consider multiple agent with $\boldsymbol{\omega}=[\omega^1, \omega^2, ..., \omega^n]$ denoting the joint policy.
Since we can adopt the framework of centralized training with decentralized execution, policies are allowed to use all agents' local observations during training. Joint empowerment $\mathcal{E}^J(\boldsymbol{a}_t, s_t)$ is computed over the joint action $\boldsymbol{a}_{t}$ and next state $s_{t+1}$ for all agents: 
\begin{equation}
    \mathcal{E}^J(s_t) = \underset{\boldsymbol{\omega}}{\max~}  \mathcal{I}
    \left[s_{t+1}, \boldsymbol{a}_t  \middle| s _t\right]
\end{equation}
This is the empowerment from all agent's actions to the state at a later point in time. It captures the causal influence of all agents actions on the overall resulting state. It also provides an upper bound on each single agent self-empowerment, a measure associated with robust single-agent behavior. It also provides a measure of the agents' collective empowerment if seen as a single organism. As such, it might also favor strategies that keep the overall population operational and increase their controllability over the world, which may positively affect the learning rate and performance.  

This reward is preferred in systems with many agents as it scales linearly with the number of agents, rather than quadratic.

\subsection{Training with Social Empowerment}
In summary, training with joint and transfer empowerment results in joint policies that are reactive, because for empowerment to be high it requires considering the decisions of others. As such, social empowerment rewards a very generalized idea of coordination that requires paying attention to each other and reliably reacting to a variation in their actions. While empowerment does not care how this reaction looks, or even if it is good, combined with the actual reward should lead to the selection of a strategy that both solves the problem, while also avoiding the brittleness that comes from not being reactive to information from other agents. Specifically, we will modify an agent's reward function $R_i$  so that it becomes:
\begin{equation}
R_{i, t}(s_{t+1}, \boldsymbol{a_t}, s_t) = r(\boldsymbol{a_t}, s_t) + \mathcal{E}(s_{t+1})
\end{equation} 
where $r$ is the environmental reward and $\mathcal{E}$ can be either transfer $\mathcal{E}^T$  or joint $\mathcal{E}^J$ empowerment. Policies $\boldsymbol{\pi}$  that are trained with $\mathcal{E}^J$  jointly select actions that result in highly empowered states (like in classical empowerment algorithms). A policy $\pi^j$  trained with $\mathcal{E}^{T, n\rightarrow j}$  selects more reactive actions, therefore also increases the empowerment of states. 


\subsection{Lower bound on $\mathcal{E}$}
Many works on empowerment operate in discrete action-state space because the computation requires integrating over all possible actions and states, which in many cases is intractable. 
We choose to operate in continuous action-state space to extend the applicability of our approach. \citet{karl2017unsupervised} introduced an efficient implementation to estimate empowerment that maximises a lower bound on  mutual information \cite{barber2003algorithm}. The lower bound on empowerment contains a source distribution, $\omega(a_t, s_t)$ and a variational approximation of the planning distribution, $q(a_t| s_{t+1}, s_t)$. By representing both distributions by a neural network, maximizing $\hat{\mathcal{I}}$ requires backpropagation using gradient \emph{ascent}. Taking inspiration from, \citet{karl2017unsupervised}, transfer empowerment can be estimated by:
\begin{equation}
\mathcal{E}^T_i(\theta, \phi) = \ln q_\xi(\boldsymbol{a}^{\omega}_{i+1}|\pi^j_\phi(s_{i+2}), s_{i+2}, s_{i+1}) - \ln \omega_\chi^k(\boldsymbol{a}^{\omega}_{i+1}|s_{i+1})
\label{eq: transfer ve}
\end{equation}where the combination of the parameters $\xi$ and $\chi$ are denoted by $\theta$.
Similarly, joint empowerment can be estimated by:
\begin{equation}
\mathcal{E}^J_i(\theta) = \ln q_\xi(\boldsymbol{a}^\omega_{i+1}|s_{i+2}, s_{i+1}) - \ln \omega_\chi(\boldsymbol{a}^\omega_{i+1}|s_{i+1})
\label{eq: joint ve}
\end{equation}
The training procedures are summarized in algorithm 1 and 2 for transfer and joint empowerment respectively. 
First, samples from the system dynamics with the behavioral policy are collected. The transition network $p$ is trained to minimize the squared error between a future state and its predicted future state. 
In algorithm 2, the estimation of empowerment is done by $\omega$, $q$ and $p$. Next, it is used as an additional reward term for the joint policy $\boldsymbol{\pi}$ to select actions that lead to high-empowered states. 
In algorithm 1, empowerment is estimated with the networks $\omega$, $q$, $p$ and $\pi^j$. So here, $\pi^j$ is trained to select actions that drive to highly empowered states and to increase the states' empowerment values. 
\algnewcommand\algorithmicforeach{\textbf{for each}}
\algdef{S}[FOR]{ForEach}[1]{\algorithmicforeach\ #1\ \algorithmicdo}
\begin{algorithm}[t]
Set random parameters for $\omega_\chi$ and $q_\xi$ with $\theta=\{\chi, \xi\}$ and $\boldsymbol{\pi}_\phi$ and $Q_\beta$.
Initialize experience replay memory with mini batches $\mathcal{B}$ with $N$ transitions.
 \begin{algorithmic}
\ForEach {$\{s_i, \boldsymbol{a}^\pi_i, s_{i+1}, r_i(\boldsymbol{a}^\pi, s_i)\} \in \mathcal{B} $}
    	\State $\epsilon_i = ( s_{i+1} - p_\psi(\boldsymbol{a}^\pi_i, s_i) )^2 $ 
    	\State $\boldsymbol{a}^\omega_{i+1} \sim \omega_\chi(s_{i+1})$, $s_{i+2} \sim p_\psi(\boldsymbol{a}^\omega_{i+1}, s_{i+1})$ 
    	\State  Estimation of $\mathcal{E}^T$ as in eq. \eqref{eq: transfer ve}
         \State  $y_i = \mathcal{E}^T_i(\theta, \phi)  + r_i + \gamma Q'_{\beta'}(s_{i+1}, \boldsymbol{\pi'}_{\phi'}(s_{i+1}))$
    \EndFor
        \State $\psi \leftarrow \psi - \lambda \nabla_\psi \frac{1}{N}\sum_i^N \epsilon_i$ 
    	\State $\theta \leftarrow \theta + \lambda \nabla_\theta \frac{1}{N}\sum_i^N\mathcal{E}^{T}_i(\theta, \phi)$
   	 	\State $\beta \leftarrow \beta + \lambda \nabla_\beta \frac{1}{N}\sum_i^N (Q_\beta(s_i, \boldsymbol{a}_i)-y_i)^2$
 		\State $\phi \leftarrow \phi + \lambda \nabla_\phi  \frac{1}{N}\sum_i^N (Q_\beta(s_i, \boldsymbol{\pi}_\phi(s_i)) + \mathcal{E}^T_i(\theta, \phi))$
 \end{algorithmic}
 \caption{$\pi^j_\phi$ trained with $\mathcal{E}^T$}
\end{algorithm}

\begin{algorithm}[t]
Set random parameters for $\omega_\chi$ and $q_\xi$ with $\theta=\{\chi, \xi\}$ and $\boldsymbol{\pi}_\phi$ and $Q_\beta$, $p_\psi$ trained as in algorithm 1.
Initialize experience replay memory with mini batches $\mathcal{B}$ with $N$ transitions.
  \begin{algorithmic}
\ForEach {$\{s_i, \boldsymbol{a}^\pi_i, s_{i+1}, r_i(\boldsymbol{a}^\pi, s_i)\} \in \mathcal{B} $}
  \State     $\boldsymbol{a}^\omega_{i+1} \sim \omega_\chi(s_{i+1})$, 
   \State  $s_{i+2} \sim p_\psi(\boldsymbol{a}^\omega_{i+1}, s_{i+1})$
    	\State  Estimation of $\mathcal{E}^J$ as in eq. \eqref{eq: joint ve}
           \State $y_i = \mathcal{E}^J_i(\theta)  + r_i + \gamma Q'_{\beta'}(s_{i+1}, \boldsymbol{\pi'_{\phi'}}(s_{i+1}))$
   \EndFor
   \State  $\theta \leftarrow \theta + \lambda \nabla_\theta \frac{1}{N}\sum_i^N \mathcal{E}^J_i(\theta)$
 \State    $\beta \leftarrow \beta + \lambda \nabla_\beta \frac{1}{N}\sum_i^N (Q_\beta(s_i, \boldsymbol{a}_i)-y_i)^2$
 \State 	$\phi \leftarrow \phi + \lambda \nabla_\phi \frac{1}{N}\sum_i^N(Q_\beta(s_i, \boldsymbol{\pi}_\phi(s_i))$
 \end{algorithmic}
 \caption{$\boldsymbol{\pi}_\phi$ trained with $\mathcal{E}^J$}
\end{algorithm}


%% file: experiments.tex
\section{Experimental Results}
\definecolor{mypink}{rgb}{0.858, 0.188, 0.478}
\subsection{Setup}
We construct three environments that test various capabilities of our approach, transfer $\mathcal{E}^T$, joint $\mathcal{E}^J$ and social influence SI \cite{jaques2019social} and baselines DDPG \cite{lillicrap2015continuous} and MADDPG \cite{lowe2017multi}. The MADDPG is a centralized version of the DDPG, meaning the value function of each agent conditions on the observations of all agents. 

We first analyze the robustness of the methods by looking at the training curves. During training, all agents change their policies and a robust learning method should deal with these perturbations. 
Second, we assess each method's ability to obtain reactive agents. We analyze this in a simulator with two agents, one sending messages and the other receiving them. The task can only be fulfilled if messages are correctly decoded, and random action taking is avoided. 
Third, we evaluate each method's ability to construct agents that remain responsive to its partners and not rely upon expectations. We hypothesize that centralized training does not incentivize this behavior. To this end, we utilize a simulator containing multiple particles that need to cover \emph{distinct} landmarks. If agents collide or go to the same landmark, we consider them not be reactive to each other.  
Last, we compare methods in a more complicated multi-agent car simulator and analyze their performance in more real-world scenarios. 

The first two scenarios are constructed by OpenAI's Multi-Agent Particle Environment \footnote{https://github.com/openai/multiagent-particle-envs}. We choose this simulator, since we are primarily interested in addressing agent interaction and are satisfied with relatively simple physics, i.e. it only contains inertia and friction. Transfer empowerment is evaluated in the first experiment and joint empowerment in the second. 
The last experiment is done in a simulator that is a combination of the multi-agent and single agent OpenAI Gym \footnote{https://github.com/openai/gym}. It allows to evaluate more complicated interactions, e.g. merging and overtaking behaviors.The code for the simulator can be found on \url{https://github.com/tessavdheiden/social_empowerment}

\begin{table*}[!htbp]
\centering
\caption{Percentage of episodes in which the listener reached the target landmark and average distance from the landmark. The right columns (3rd-6th) show results of the task in which the environment contains less channels than landmarks (L=6, C=5) and obstacles (O=6). Alternating frequency is the rate at which the speaker alters its message per episode. A low number indicates that the speaker efficiently uses its communication channel. }
\begin{tabular}{l*7c}
\toprule
    &  \multicolumn{2}{c}{$L=3, C=3, O=0$} & \multicolumn{5}{c}{$L=6, C=5, O = 6$}\\
{}              & \textbf{Reward}& \textbf{Average}  & \textbf{Target }      & \textbf{Reward}    & \textbf{Average} & \textbf{Target} & \textbf{Obstacle }\\
{}              & \textbf{} 	& \textbf{distance}  & \textbf{reach \%}    & \textbf{}          & \textbf{distance} & \textbf{reach \%} & \textbf{hit \% } \\
\midrule
\textbf{DDPG} & -0.439     & 0.456               &  32.0                   & -0.851           & 0.653             & \cellcolor{lgrey}30.1      & 60.8            \\
\textbf{MADDPG} &\cellcolor{lgrey} -0.301  & \cellcolor{lgrey} 0.133 &\cellcolor{grey} 92.0 &\cellcolor{lgrey} -0.596& \cellcolor{lgrey}0.260     & 40.5                 & \cellcolor{lgrey} 49.4   \\
{SI}       		& -0.569       & 0.290                &  88.0                  & -0.666            & 0.360 & 31.5                 & 51.2               \\
{$\mathcal{E}^{T}$} &\cellcolor{grey}-0.292    & \cellcolor{grey}0.120  & \cellcolor{lgrey} 91.1 &\cellcolor{grey}-0.422&   \cellcolor{grey}0.072& \cellcolor{grey}61.1 & \cellcolor{grey} 31.1 \\
\bottomrule
\end{tabular}
\label{tab:comm}
\end{table*}

\begin{table*}[!htbp]
\centering
\caption{Average percentage of collisions and successes per episode consisting of 25 time steps. Distance from landmark are averaged over all time steps. An episode is successful if all agents are within .1 distance to a distinct landmarks at the last 5 time steps. Shaded areas indicate best (dark grey) and second best (light grey) numbers. }
\label{tab: navigation}
\begin{tabular}{l*5c}
\toprule
{}  &  \textbf{Reward}  & \textbf{Average distance}   & \textbf{Collisions \%} &  \textbf{Success \%}    \\
\midrule
\textbf{DDPG}   				& -2.302       								& 1.324 									&  50.0				& 69.3  \\
\textbf{MADDPG}   		& -2.198   							& 1.108         								&  53.3              	& 80.5     \\
\textbf{SI}   						& -2.461    						& 1.459  									&   70.3   					& 61.9  \\
\textbf{$\mathcal{E}^T$}   &\cellcolor{lgrey} -2.098    	&  \cellcolor{lgrey}0.857  		&  \cellcolor{lgrey} 43.3   						& \cellcolor{lgrey}89.9  \\
\textbf{$\mathcal{E}^J$}   &\cellcolor{grey} -2.063    	&  \cellcolor{grey}0.957  		&  \cellcolor{grey} 13.3   						& \cellcolor{grey}95.9  \\
\bottomrule
\end{tabular}
\end{table*}

\subsection{Training with Empowerment}
During training, all agents change their policies, and a robust learning method should deal with these perturbations. We hypothesize that training with empowerment results in robust methods because it incentivizes being responsive to changes in others' behaviors. 

\begin{figure}
     \centering
     \textbf{Training Curves}\par\medskip
     \begin{subfigure}[b]{0.2\textwidth}
         \centering
         \includegraphics[width=\textwidth]{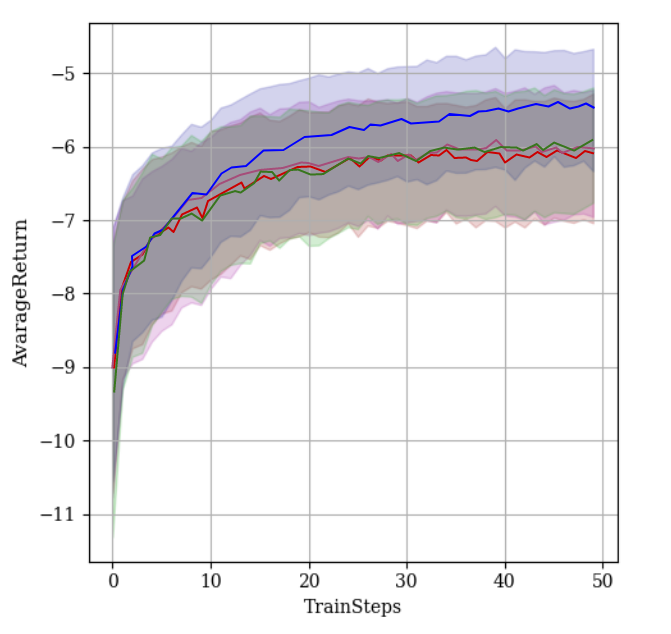}
         \caption{Cover Landmarks}
         \label{fig:train1}
     \end{subfigure}
    \begin{subfigure}[b]{0.2\textwidth}
         \centering
         \includegraphics[width=\textwidth]{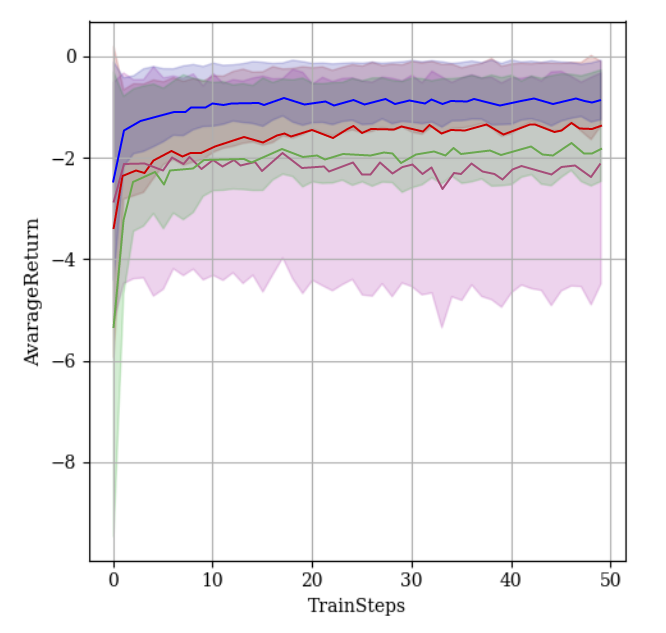}
         \caption{Speaker Listener}
         \label{fig:train2}
     \end{subfigure}
        \caption{Training curves for Cover Landmarks and Speaker Listener tasks.  \textcolor{green}{---} DDPG \textcolor{red}{---} MADDPG \textcolor{mypink}{---} SI \textcolor{blue}{---} $\mathcal{E}$  }
        \label{fig:two graphs}
\end{figure}

Figures \ref{fig:train1} and \ref{fig:train2} show the training curves for the baselines and our method in the first two tasks. When an agent has empowerment as an additional utility function the value of the average return is higher. 

\subsection{Experiment \Romannum{1}: Cooperative Communication}
The first task involves one silent moving agent ($j$) and the other one ($k$) sending messages that help navigate the listening agent. The objective for the moving agent is to reach a target landmark out of $L$ landmarks by decoding the speaker's message (a symbol chosen from $C$ distinct symbols).  The team reward is the Euclidean distance from the target landmark. The task requires both agents to reliably act and pay attention to each other's actions. The agents can fail if the speaker outputs random symbols or if the listener incorrectly decodes or ignores the messages. 

If $L=C$, the speaker simply needs to output a unique symbol identifying the target landmark. By having more landmarks than distinct symbols, i.e. $L > C$, the speaker is forced to send more indicative signals, e.g. indicating movement direction. Last, by adding $O$ obstacles that are \emph{only visible to the speaker}, the listener is challenged more to decode the messages correctly and navigate precisely. 

Table \ref{tab:comm} shows the performance of the baselines and our method on the two versions of this task. We present  $\mathcal{E}^{T, k\rightarrow j}$ which should motivate the listener ($j$) to adapt its policy to be more responsive to the messages. $\mathcal{E}^{T, j\rightarrow k}$ should have the speaker adapt its policy. However, as the messages are not blocked, there is little in term of policy change for the speaker that is helpful to increase the empowerment. 

The lowest number of obstacle hits was obtained by training with $\mathcal{E}^{T, k\rightarrow j}$. This can be explained by the listener's policy adaptation to decode the messages and navigate better.

\begin{figure}
     \centering
     \textbf{Speaker Listener Task}\par\medskip
     \begin{subfigure}[b]{0.15\textwidth}
         \centering
         \includegraphics[width=\textwidth]{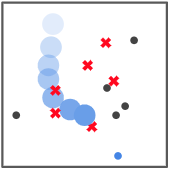}
         \caption{DDPG}
         \label{fig:ddpg sl}
     \end{subfigure}
     \hfill
     \begin{subfigure}[b]{0.15\textwidth}
         \centering
         \includegraphics[width=\textwidth]{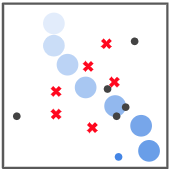}
         \caption{MADDPG}
         \label{fig:maddpg sl}
     \end{subfigure}
     \hfill
     \begin{subfigure}[b]{0.15\textwidth}
         \centering
         \includegraphics[width=\textwidth]{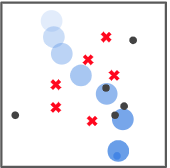}
         \caption{$\mathcal{E}^{T, k\rightarrow j}$}
         \label{fig:maddpg_e sl}
     \end{subfigure}
        \caption{The listener (big blue circle) has to move to a target landmark (blue small circle) out of $L=6$ landmarks (black small circles) while avoiding the obstacles (red crosses). The sender can only send messages of length $C=5 < L$. }
        \label{fig:three graphs comm}
\end{figure}

Fig. \ref{fig:three graphs comm} shows an episode. DDPG does not reach the target and hits obstacles, see \ref{fig:ddpg sl} . While MADDPG avoids the obstacles (see, \ref{fig:maddpg sl}), it remains at a larger distance from the target than our method, in \ref{fig:maddpg_e sl}. 

\begin{table*}[!htbp]
\centering
\caption{Three cars need to avoid collisions and obstacles on the road. All agents are rewarded for staying on the road and penalized if they hit an obstacle or another car. }
\begin{tabular}{l*4c}
\toprule
 &  \multicolumn{2}{c}{Junctions} & \multicolumn{2}{c}{Obstacles}\\
 & \textbf{Off road \%} &  \textbf{Collisions \%}  &  \textbf{Obstacle hit \%} &  \textbf{Collisions \%} \\
\midrule
\textbf{MADDPG}    &  11.8  & 12.7   & 29.3  & 31.9\\
\textbf{$\mathcal{E}^J$}  &  \cellcolor{grey}6.6   &  \cellcolor{grey}5.6   & \cellcolor{grey}15.9  & \cellcolor{grey}12.4\\
\bottomrule
\end{tabular}
\label{tab:car}
\end{table*}

\subsection{Experiment \Romannum{2}:  Cooperative Navigation}
In this task three agents have to move to three distinct landmarks. They are not told to go to a specific one, but rather need to attend and react to their partners and choose distinct ones. They observe the relative coordinates to their partners and landmarks. The team reward is the Euclidean distance between a landmark and the closed agent, summed over all landmarks and agents obtain a penalty for colliding with each other. If agents collide or go to the same landmark, we consider them not to be reactive to each other. We hypothesize that joint empowerment has agents moving to distinct landmarks faster without colliding and thus obtain a higher reward. 

Table \ref{tab: navigation} presents results. 
Our method achieves the lowest average distance and the lowest number of collisions. Near collisions are avoided by using joint empowerment as additional reward, because it is low in states in which the agents have low maneuverability. 

\begin{figure}[!htbp]
     \centering
     \textbf{Cover Landmarks Task}\par\medskip
     \begin{subfigure}[b]{0.15\textwidth}
         \centering
         \includegraphics[width=\textwidth]{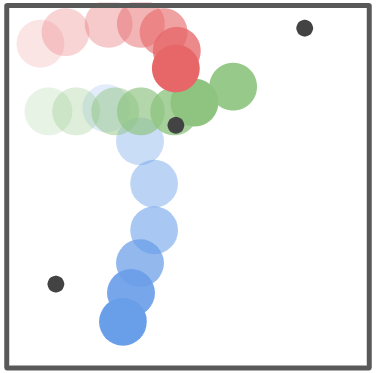}
         \caption{DDPG}
         \label{fig:y ddpg nav}
     \end{subfigure}
     \hfill
     \begin{subfigure}[b]{0.15\textwidth}
         \centering
         \includegraphics[width=\textwidth]{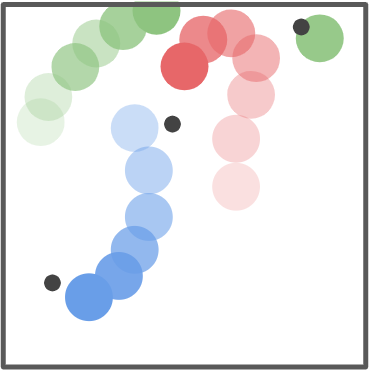}
         \caption{MADDPG}
         \label{fig:maddpg nav}
     \end{subfigure}
     \hfill
     \begin{subfigure}[b]{0.15\textwidth}
         \centering
         \includegraphics[width=\textwidth]{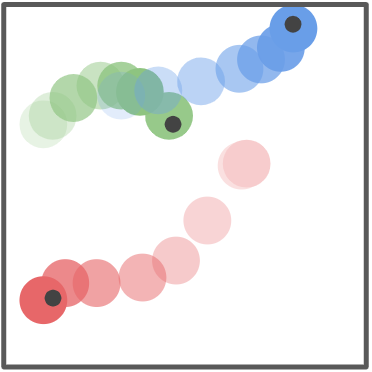}
         \caption{$\mathcal{E}^J$}
         \label{fig:maddpge nav}
     \end{subfigure}
        \caption{The task of the agents (big circles) is to reach distinct landmarks (small circles). Each 3rd time step is shown (from transparent to opaque).}
        \label{fig:three graphs nav}
\end{figure}

Fig. \ref{fig:three graphs nav} shows a full episode. The three agents are the big colored circles and the landmarks are the smaller black ones. MADDPG and MADDPG+$\mathcal{E}$ reach distinct landmarks. Empowerment captures the causal influence of all agents' actions on the overall  resulting state and agents avoid states with limited maneuverability. This can be seen by the blue agent that does not move to the closest landmark, but rather one that is further away from its green partner \subref{fig:maddpge nav}. MADDPG has two colliding agents. See Fig. \subref{fig:maddpg nav}.

\subsection{Experiment \Romannum{3}: Cooperative Driving}
In the last task we consider three autonomous vehicles that have to interact in novel
traffic situations. The simulator contains physical properties such as road friction and inertia. Agents observe a stack of 4 top view grey scale images, centered around their center of mass from which they need to infer the relative location of their partners and obstacles, and their own speed. They are rewarded by the number of pixels in their observation corresponding to the color of the road, motivating them to stay on the road, and are penalized for colliding with each other. 

Intersections and obstacles are added to assess the agents' abilities to react in time to others. For example, if multiple agents enter a point where two roads meet or a location with many obstacles, agents can collide if neither one gives way. We therefore count the number of collisions and obstacle hits.

Table \ref{tab:car} shows the two different scenarios, the road with junctions and obstacles. Agents trained with empowerment results in agents outperforming the MADDPG on all metrics. 

\begin{figure}[!htbp]
     \centering
    \textbf{Cooperative Driving}\par\medskip
     \begin{subfigure}[b]{0.11\textwidth}
         \centering
         \includegraphics[width=\textwidth]{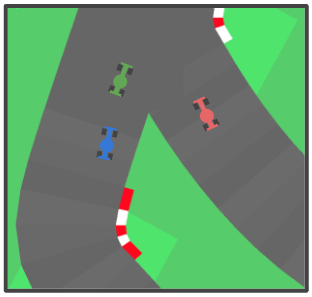}
         \caption{MADDPG}
         \label{fig:merge no e}
     \end{subfigure}
     \begin{subfigure}[b]{0.11\textwidth}
         \centering
         \includegraphics[width=\textwidth]{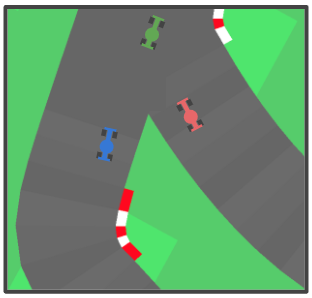}
         \caption{$\mathcal{E}^J$}
         \label{fig:merge e}
     \end{subfigure}
     \begin{subfigure}[b]{0.11\textwidth}
         \centering
         \includegraphics[width=\textwidth]{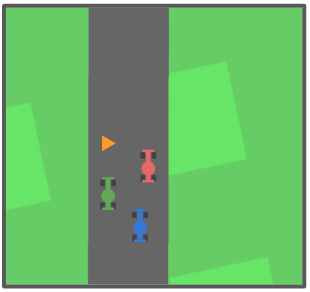}
         \caption{MADDPG}
         \label{fig:highway no e}
     \end{subfigure}
          \begin{subfigure}[b]{0.11\textwidth}
         \centering
         \includegraphics[width=\textwidth]{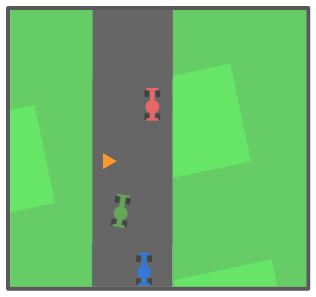}
         \caption{$\mathcal{E}^J$}
         \label{fig:highway e}
     \end{subfigure}
        \caption{Comparison in the dynamic car simulator in which 3 vehicles need to merge \subref{fig:merge no e}, \subref{fig:merge e} and avoid obstacles \subref{fig:highway no e}, \subref{fig:highway e} while staying on the road and keeping safe distance between each other.  }
        \label{fig:three graphs car}
\end{figure}

In Fig \ref{fig:three graphs car} we show two snapshots in the two scenarios. The agents that are trained with the MADDPG come closer to each other and respond very late. In contrast, when using $\mathcal{E}^J$ the vehicles respond quicker and give others space to merge in or to avoid obstacles. Agents (blue car) give way, see \subref{fig:merge e} and \subref{fig:highway e}, to let the other cars (green, red) pass.

\subsection*{Technical details}
All methods are trained for 25k episodes, each episode consisting of 25 time-steps in the first two task and 50 time-steps in the last. The hyper-parameters as well as the optimization algorithm are equal to \cite{lowe2017multi}. In the last experiment agents obtained a stack of 4 images and all networks had a 5-layered CNN to convert the stack into a single-dimension vector.  The code for the implementation can be found on \url{https://github.com/tessavdheiden/social_empowerment}

%% file: conclusion.tex
\section{Conclusions and Future Work}
In this paper, we consider the problem of agents that overfit to their training partners and that adapt strategies that act under certain expectations, rather than react to their actions. We propose to use social empowerment that biases the learning process to construct agents that remain reactive to their partners' behaviors. 

In three experiments we show how empowerment motivates the construction of agents that are responsive to each other's behavior. We have explained different variants, transfer empowerment that can be applied between a pair of agents, and a faster computational variant, joint empowerment, which can be applied for a group of agents with any size. 

Our method, uses centralized training with decentralized execution, thus can be deployed with local partial observations of the policies. We show that not only agents coordinate better by taking into account the dynamics of the whole group, but also more precisely, avoiding dangerous situations.  

An exciting direction of research, which we have not explored is cooperation with partners that agents have not seen during training, e.g. humans. Furthermore, many real-life scenarios have competitive aspects, and agents can also be trained by minimizing the empowerment of their opponents. Focusing on the before mentioned problems allows tackling a large number of scenarios, such as robots developed by different brands that need to interact with humans.